\documentclass[aps,pra,twocolumn,10pt,noshowpacs]{revtex4-1}
\usepackage{hyperref}
\usepackage{amssymb}
\usepackage{microtype}
\usepackage{graphicx}
\usepackage{amsmath}
\usepackage{amsfonts}
\usepackage{dcolumn}
\usepackage{bm}
\usepackage{color,xspace}
\usepackage{calc}
\usepackage{ifthen}
\usepackage{xfrac}
\usepackage{booktabs}
\usepackage{float}
\usepackage{xspace}

\graphicspath{{./figures/}}

\newcolumntype{d}[1]{D..{#1}}
\newcommand\mc[1]{\multicolumn{1}{c}{#1}}

\newlength{\blength}
\newcommand{\betai}{\beta_{\text{i}}}
\newcommand{\betacGS}{\beta_{\text{GS}}}
\newcommand{\betacGSV}{\beta_{\text{GS}}^{\text{V}}}
\newcommand{\betacInf}{\beta_{(\infty)}}
\newcommand{\betacInfV}{\beta_{(\infty)}^{\text{V}}}
\newcommand{\betacOne}{\beta_{(1)}}
\newcommand{\betacTwo}{\beta_{(2)}}
\newcommand{\dblqt}[1]{\textquotedblleft #1\textquotedblright\xspace}
\newcommand{\rcolh}{\,\mathord{:}\,}
\newcommand{\rcol}{\hspace{-0.065em}\rcolh}
\newcommand{\wdth}{\mathit{w}}
\DeclareMathOperator{\sech}{sech}
\begin{document}

\title{Metastability versus collapse following a quench in attractive
Bose-Einstein condensates}

\author{Jake Golde}
\affiliation{Department of Physics, University of Massachusetts Boston, Boston, MA 02125, USA}
\author{Joanna Ruhl}
\affiliation{Department of Physics, University of Massachusetts Boston, Boston, MA 02125, USA}
\author{Sumita Datta}
\affiliation{Department of Physics, The ICFAI Foundation for Higher Education, Hyderabad - 501203, India}
\author{Boris A. Malomed}
\affiliation{Department of Physical Electronics, School of Electrical Engineering,
Faculty of Engineering, Tel Aviv University, Tel Aviv 69978, Israel}
\affiliation{Laboratory of Nonlinear-Optical Informatics, ITMO University, St. Petersburg 197101, Russia}
\author{Maxim Olshanii}
\email{Maxim.Olchanyi@umb.edu}
\affiliation{Department of Physics, University of Massachusetts Boston, Boston, MA 02125, USA}

\author{Vanja Dunjko}
\email{Vanja.Dunjko@umb.edu}
\affiliation{Department of Physics, University of Massachusetts Boston, Boston, MA 02125, USA}

\date{\today}

\begin{abstract}
We consider a Bose-Einstein condensate (BEC)\ with attractive two-body interactions in a cigar-shaped trap, initially prepared in its ground state for a given negative scattering length, which is quenched to a larger absolute value of the scattering length. Using the mean-field approximation, we compute numerically, for an experimentally relevant range of aspect ratios and initial strengths of the coupling, two critical values of quench: one corresponds to the weakest attraction strength the quench to which causes the system to collapse before completing even a single return from the narrow configuration (\dblqt{pericenter}) in its breathing cycle. The other is a similar critical point for the occurrence of collapse before completing \emph{two} returns. In the latter case, we also compute the limiting value, as we keep increasing the strength of the post-quench attraction towards its critical value, of the time interval between the first two pericenters. We also use a Gaussian variational model to estimate the critical quenched attraction strength below which the system is stable against the collapse for long times. These time intervals and critical attraction strengths---apart from being fundamental properties of nonlinear dynamics of self-attractive BECs---may provide clues to the design of upcoming experiments that are trying to create robust BEC breathers.
\end{abstract}

\maketitle


\section{Introduction \label{sec:introduction}}

\begin{figure}[tbp]
\begin{center}
\includegraphics[width=0.4%
\textwidth,keepaspectratio=true,draft=false,clip=true]{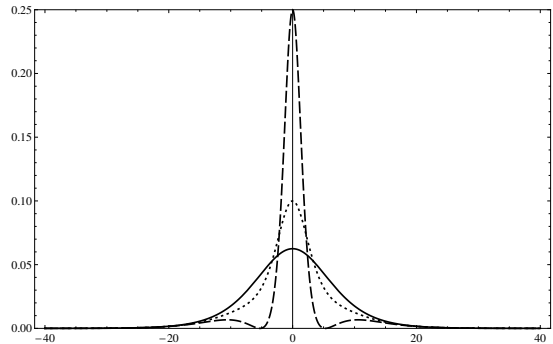}
\end{center}
\caption{ \textbf{The $\mathbf{1\rcol 3}$ breather.} The density profile, at three points of
time, of an exact two-soliton solution of the integrable 1D NLSE, built of
fundamental solitons with coinciding centers, zero velocities, and norms $%
1/4 $ and $3/4$. The solid line corresponds to the initial profile, the
dotted line to the profile at $1/4$ density period, and the
dashed line to $1/2$ period. }
\label{basic_breather}
\end{figure}

Presently, at least two experimental groups \cite%
{Everitt2015_1509.06844,Hulet2016_private} are trying to realize
Gross-Pitaevskii breathers \cite%
{zakharov1972_118,Satsuma1974_284,Sakaguchi2004_066613,prilepsky2007_036616,Yanay2009_033145,Dunjko2015_1501.00075}
in attractive Bose-Einstein condensates (BECs) \cite{Pethick2008}. Their
immediate goal is to create conditions such that the system is
well-described by the one-dimensional (1D) integrable nonlinear Schr{\"{o}}%
dinger equation (NLSE),
\begin{equation}
i\psi _{t}=-\frac{1}{2}\psi _{xx}+\eta \left\vert \psi \right\vert ^{2}\psi
\label{NLSE}
\end{equation}%
\cite{ablowitz1981}, with $\eta <0$, and then to excite a \emph{breather}%
---a \dblqt{nonlinear superposition} of fundamental NLSE
solitons whose centers of mass coincide and which are at rest relative to
each other \cite{Cardoso2010_2640}. Experimentally, the effectively 1D
approximation is provided by placing the BEC in a elongated, cigar-shaped
trap \cite{castin2004_89}. The breather that is the principal target of the
current experimental efforts is shown in Fig.~\ref{basic_breather}. Apart
from being objects of interest in their own right, breathers are also
potentially useful in atomic interferometry \cite{Dunjko2015_1501.00075}.

Breathers may be excited by quenching the nonlinearity strength $\eta $ in
Eq. (\ref{NLSE}). Here and below, by the \dblqt{strength} of a
negative quantity we mean its magnitude, and, accordingly, by
\dblqt{stronger} and \dblqt{weaker} we will mean
interaction strengths which are larger and smaller in magnitude,
respectively. In BECs, the quench can be applied via magnetically
tuned Feshbach resonances \cite{Chin2010_1225,Nguyen918_2014}. Here one may take advantage of a special
property of the integrable NLSE: Consider an $N$-soliton breather, composed
of $N$ fundamental solitons with norm ratios $1 \rcol 3 \rcolh 5 \rcolh 7 \rcolh 
\ldots\,\rcolh 
(2N-1)$
(an \dblqt{odd-norm-ratio breather}), with their phases
synchronized initially. At the point in its breathing cycle at which it is
the widest (the \dblqt{apocenter}), its wave function is the same
as for a \emph{fundamental} soliton generated by Eq. (\ref{NLSE}) with the
value of $\eta $ which is smaller by a factor of $N^{2}$ \cite%
{Satsuma1974_284}. Thus, for example, to excite the $1 \rcol 3$ breather, we first prepare a fundamental soliton of the NLSE, and then apply the quench,
suddenly multiplying $\eta $ by $4$.

\begin{figure}[tbp]
\begin{center}
\includegraphics[width=0.4%
\textwidth,keepaspectratio=true,draft=false,clip=true]{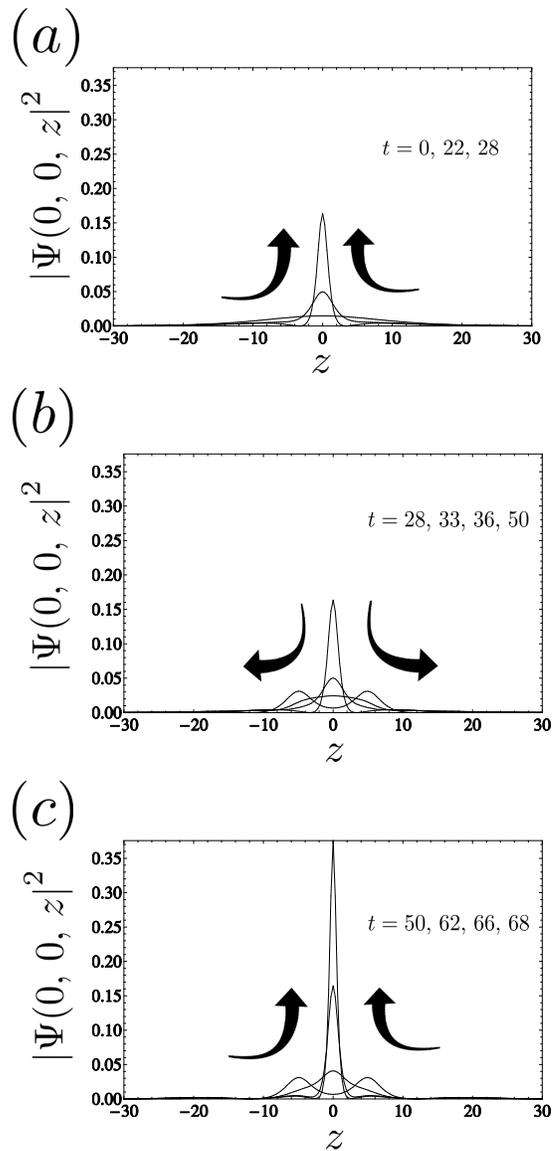}
\end{center}
\par
\caption{ \textbf{A complex breather.} The axial density profile at several
points in time of a 3D breather. We are using the \dblqt{natural
units} introduced in the text preceding Eq.~(\protect\ref%
{GPELabGPE}). The breather is obtained by initializing the system in the
ground state of the 3D GPE with $\protect\gamma _{z}=0$ and $\protect\beta %
=(1/8)\protect\beta _{\text{GS}}$; here $\protect\beta _{\text{GS}}$ is the
ground-state critical value of the coupling constant. Then, at $t=0$, we
apply the quench factor $9.6$ to produce $\protect\beta =1.2\protect\beta _{%
\text{GS}}$. Because the quench is close to $3^{2}$, the breathings resemble
those of the exact $3$-soliton solution in 1D. Note that the post-quench
coupling strength is supercritical, and the variational model of Sec.~%
\protect\ref{sec:variational} suggests that the system eventually collapses.
Panel \textbf{(a)} shows the solitary wave initially narrowing until it
reaches the \dblqt{pericenter} at $t=28$. After that, as shown in
\textbf{(b)}, the system widens until reaching the \dblqt{apocenter} at $t=50$. The double-peak structure at this moment is a notable
feature of the 1D $3$-soliton, although it can also occur for $2$-solitons
whose constituent fundamental solitons have comparable norms \protect\cite%
{Dunjko2015_1501.00075}. In panel \textbf{(c)} the system narrows again,
until reaching a new, even narrower \dblqt{pericenter}. Then, the
breather begins to widen again (not shown here). Additionally, although it
is not well visible on the scale of these plots, following the quench there
appears radiation emitted by the breather. Radiation is present in the
integrable solution as well, if the quench factor is not exactly the square
of an integer \protect\cite{Satsuma1974_284}. }
\label{complex_breather}
\end{figure}

Following the quench, the system will start to \dblqt{breathe,} alternating
between a narrow density profile (the \dblqt{pericenter}) and a
wide one (the \dblqt{apocenter}). For the quench factor close to $%
4 $, numerical simulations produce breathings of the longitudinal density
profile which resemble the picture generated by the exact two-soliton
solution displayed in Fig.~\ref{basic_breather}, which includes the
development, at the pericenter, of the two \dblqt{dwarf satellite
peaks} adjacent to the central peak. Note that the satellite peaks are an
interference effect, and do not imply fragmentation of the matter wave.
Generally, this picture is \emph{quantitatively} correct when (1) the 1D
approximation is accurate enough, for which we need the \emph{effective
coupling strength}---the product of the absolute value of the scattering length and of the number of particles---to be as small as
possible; (2) there is no longitudinal confinement; and (3) the quench
factor is close to $4$. For smaller quenches, the satellite peaks do not
emerge, while the profile (qualitatively speaking) oscillates between the
solid-line and the dotted-line shapes in Fig.~\ref{basic_breather}. On the
other hand, for much larger quenches, the breathings become more complex. In
particular, if the quench is close to $n^{2}$, with integer $n$, the
breathings resemble the well-known exact solutions for $n$-solitons \cite%
{Satsuma1974_284}. For example, the three-dimensional (3D) breathings
displayed in Fig.~\ref{complex_breather} are similar to those of the
integrable 1D $3$-soliton with the norm ratio $1\rcol 3\rcolh 5$, since the
quench is close to 9.

Actual experiments deal with 3D settings, even if they correspond to
elongated traps. As is well known, 3D self-attractive BECs are unstable
against collapse \cite%
{Zakharov1975_465,Zakharov1985_154,Zakharov1986_773,sulem1999,Berge2002_136}
whenever the coupling strength exceeds a certain critical value (however,
see \footnote{%
In principle, for sufficiently long times, many-body quantum effects cause
the system to collapse for any positive value of the coupling, via
macroscopic quantum tunneling (across the kinetic-energy barrier that makes
the system stable at the mean-field level) \cite%
{Ruprecht1995_4704,Kagan1996_2670,Kagan1997_2604,Kagan1998_933,Eleftheriou2000_043601}%
. However, for a fixed coupling constant, the tunneling rate is suppressed,
at least exponentially, by the difference between the actual number of atoms
and the critical one; hence it is negligible, except for very close to the
critical point \cite{Ueda1999_3317}.}). The collapse has been observed and
extensively studied experimentally too \cite%
{Gerton2000_692,Roberts2001_4211,Donley2001_295,Cornish2006_170401,Altin2011_033632,Compton2012_063601,Eigen2016_041058}%
. On the theoretical side, a study of direct relevance to the present work
was reported in Ref. \cite{gammal2001_055602}. Similar to what we do below,
that work also used the mean-field approximation, i.e., the Gross-Pitaevskii
equation [GPE, see Eq.~(\ref{3DGPE}) below]. It aimed to compute the
\dblqt{ground-state (GS) critical value} (i.e., the value of the
coupling strength below which there is a stable GS within the mean-field
model, whereas the GS does not exist above the critical value) for the full
range of aspect ratios of the cylindrically symmetric trap. For
completeness' sake, we reproduce those results in Table~\ref{GammalValues}
below.

However, breathers are excited states of the system, rather than a GS. Thus,
the boundaries of the parameter regime in which the system is stable against
dynamical collapse of such states should be re-investigated. To this end,
suppose we prepare the BEC in its GS at some sub-critical value of the
coupling constant, $\betai$ [see Eq.~(\ref{3DGPE}) below
and the related text for the precise definition of $\beta $], and then quench it
to some $\beta _{q}=q\betai$, with $q>1$. For every given aspect
ratio of the trap and $\betai$, there is a critical value of $%
\beta _{q}$, $\beta _{(\infty )}$, such that the system is stable against
collapse at $\left\vert \beta _{q}\right\vert <\left\vert \beta _{(\infty
)}\right\vert $ (at the GPE\ level) for indefinitely long times, whereas the
collapse eventually happens at $\left\vert \beta _{q}\right\vert >\left\vert
\beta _{(\infty )}\right\vert $. Intuitively, one may expect that $%
\left\vert \beta _{(\infty )}\right\vert <\left\vert \beta _{\text{GS}%
}\right\vert $, where $\beta _{\text{GS}}$ is the GS critical value of the
coupling. The variational analysis reported in Sec.~\ref{sec:variational}
turns the intuition into an explicit argument, and provide an estimate for $%
\beta _{(\infty )}$ (whose absolute value indeed turns out to be bounded by $%
\left\vert \beta _{\text{GS}}\right\vert $, within the variational model).
Nevertheless, it is not known, and we do not aim to address this rather
complex issue here, whether $\left\vert \beta _{(\infty )}\right\vert
<\left\vert \beta _{\text{GS}}\right\vert $ always holds in the framework of
the full GPE.

Note that experimentally relevant time scales cannot be effectively infinite
relative to the breathing period \cite{Hulet2016_private}. Thus, it is
relevant to consider metastability too.

What we observe numerically is that, in some parameter regions, immediately following the quench, the system keeps shrinking while the height of the
central peak increases, leading to an \emph{immediate collapse}. However, at
other values of the parameters, the system completes one or more cycles of
sequential narrowing and broadening before it finally collapses, which may
be categorized as a \emph{delayed collapse} \cite{Biasi2017_032216}, a kind of metastability. The variational treatment produced in Sec.~\ref{sec:variational} offers a qualitative picture for this kind of the metastability.

We should mention that Biasi et al. \cite{Biasi2017_032216} carried out a study very much related to ours, which we discuss in Appendix~\ref{sec:Biasi}. We should also mention the work of Mardonov et al. \cite{Mardonov2015_043604}, which studies how one can use spin-orbit coupling to control the collapse of a 2D condensate, even blocking it completely in some regimes.

We will study two kinds of quench-critical coupling strengths, which can be
done with reasonable accuracy. One is $\beta _{(1)}$, the smallest (in its
absolute value) coupling strength the quench to which causes the system to
collapse immediately, before completing even \emph{one} return from the
pericenter. The other, $\beta _{(2)}$, is a critical value for the
occurrence of the collapse before completing \emph{two} returns. In the
latter case, we also compute the limiting value, as we keep increasing the
coupling strength towards the critical point, of the time interval between
the first two pericenters. Here we assume (as corroborated by numerical
results) that, as we keep all other parameters fixed, if the system does not
collapse before completing a single return from the pericenter for a
particular value of the post-quench coupling, it will not collapse either
for any smaller value. Conversely, if the system does collapse for a given
value of the coupling, it will collapse as well for all larger values.
Similar statements hold when considering the collapse before completing two
returns. This justifies defining the term \dblqt{critical} for
the above-mentioned special values of the quench factor.

As usual, we start the analysis from the 3D GPE including the
harmonic-oscillator trapping potential with transverse and longitudinal
frequencies $\omega _{\perp }$ and $\omega _{z}$ \cite{Pethick2008},
\begin{gather}
i\hbar \partial _{t}\Psi =-\frac{\hbar ^{2}}{2m}\nabla ^{2}\Psi + g N_{\text{a}%
}\left\vert \Psi \right\vert ^{2}\Psi  \notag \\
+\frac{1}{2}m\left( \omega_{r}^{2}\left( x^{2}+y^{2}\right) +\omega_{z}^{2}z^{2}\right) \Psi \,,  \label{3DGPE}
\end{gather}%
with normalization $\int_{V}\,\left\vert \Psi (\vec{r},\,t)\right\vert
^{2}\,dV=1$. Here $m$ is the atomic mass, $N_{\text{a}}$ the number of
atoms, and $g=4\pi \hbar ^{2}a/m$ the coupling constant, $a$ being the
scattering length of interatomic interactions. Dimensional analysis shows that the problem has two dimensionless parameters. The usual choice are the following two \cite{gammal2001_055602}: the anisotropy parameter $\gamma _{z}\equiv \omega _{z}/\omega _{r}$, and the parameter $\beta \equiv a_{3D}N_{\textrm{a}}/a_{\perp}$, where $a_{\perp }=\sqrt{\hbar /(m\omega _{r})}$, which compares the strength of the nonlinearity to the strength of the transversal confinement.

For numerical work, it makes sense to use \dblqt{natural} units, namely those in which $\hbar=m=\omega_{r}=1$ (see Appendix~\ref{app:natural}).

Note that for realistic description of experiments, one would need to include the three-body losses in the model as well as gain from the thermal cloud. But for the purposes of this paper, we will take a more theoretical approach and study the GPE without these terms.

In Table~\ref{GammalValues}, for the sake of completeness we include the tabulated values of $\left|\betacGS(\gamma_{z})\right|$ from Gammal et al. \cite{gammal2001_055602}.

\begin{table}[ht]
\begin{tabular}{c@{\hspace{2ex}}  c c c c c c c c }
\toprule
$\gamma_{z}$ & 0.01 & 0.02 & 0.05 & 0.1 & 0.2 & 0.3 & 0.5 & 1.0\\
\hline
\mbox{}\vspace{\blength}\mbox{} & & & & & \\
$\left|\betacGS(\gamma_{z})\right|$ & 0.676 & 0.676 & 0.677 & 0.675 & 0.666 & 0.654 & 0.629 & 0.575
\\ \bottomrule
\end{tabular}
\caption{The values of $\left|\betacGS(\gamma_{z})\right|=(N_{\textrm{a}}\left|a_{3D}\right|)_{\text{crit}}/a_{\perp}$ as a function of the aspect
ratio $\protect\gamma _{z}$, for cigar-shaped trapping configurations, as
reproduced from Ref. \protect\cite{gammal2001_055602}. In terms of notation
adopted in that work, $\protect\gamma _{z}=\protect\lambda $ and $\left|\betacGS(\gamma_{z})\right|=\lambda^{-1/6}k(\lambda)$.}
\label{GammalValues}
\end{table}

\section{The variational model \label{sec:variational}}

As the first step of the analysis, we address the problem variationally \cite{Anderson1979_1838,Anderson1983_3135,Desaix1989_2441,Desaix1991_2082,Perez-Garcia1996_5320}. We will be using the following effective equations for matter-wave Gaussian widths ($\wdth_{x}$, $\wdth_{y}$, and $\wdth_{z}$) derived in Ref.
\cite{Perez-Garcia1996_5320}; the approach of Ref. \cite%
{Eleftheriou2000_043601}, although not variational, is similar in spirit. Very similar equations were also used in the study of collapse in Ref.~\cite{Carr2002_063602}, except that here it was assumed that the density profile in the $z$-direction is a $\sech$-squared, the shape of the exact 1D single soliton.
\begin{align}
m\ddot{\wdth}_{x}& =\frac{\hbar ^{2}}{m\wdth_{x}^{3}}+\frac{gN_{\text{a}}}{(2\pi
)^{3/2}}\frac{1}{\wdth_{x}^{2}\wdth_{y}\wdth_{z}}-m\omega _{r}^{2}\wdth_{x}\,,  \notag \\
m\ddot{\wdth}_{y}& =\frac{\hbar ^{2}}{m\wdth_{y}^{3}}+\frac{gN_{\text{a}}}{(2\pi
)^{3/2}}\frac{1}{\wdth_{x}\wdth_{y}^{2}\wdth_{z}}-m\omega _{r}^{2}\wdth_{y}\,,  \label{VA} \\
m\ddot{\wdth}_{z}& =\frac{\hbar ^{2}}{m\wdth_{z}^{3}}+\frac{gN_{\text{a}}}{(2\pi
)^{3/2}}\frac{1}{\wdth_{x}\wdth_{y}\wdth_{z}^{2}}-m\omega _{z}^{2}\wdth_{z}\,,  \notag
\end{align}%
where $\ddot{\wdth}_{x}\equiv d^{2}\wdth_{x}/dt^{2}$, and similarly for $\wdth_{y}$ and $\wdth_{z}$. Here the Gaussian widths
are variational parameters in the Gaussian ansatz,
\begin{gather}
\Psi (x,\,y,\,z,\,t)=A(t)e^{i\theta (t)}  \notag \\
\times \exp \left[ -\frac{1}{2}\left( \frac{x^{2}}{\wdth_{x}^{2}(t)}+\frac{y^{2}%
}{\wdth_{y}^{2}(t)}+\frac{z^{2}}{\wdth_{z}^{2}(t)}\right) \right.\notag \\
\left.
+\frac{i}{2\hbar }\left(
p_{x}(t)x^{2}+p_{y}(t)y^{2}+p_{z}(t)z^{2}\right) \right] \,.  \label{ans}
\end{gather}%
The amplitude in the ansatz may be expressed in
terms of the Gaussian widths and the norm $N(t)$ of the wave function (the norm is not a priori constrained to be constant, though we do assume that the wave function is normalized to unity at $t=0$, $N(0)=1$):
\begin{equation*}
A(t)=\frac{1}{\pi ^{3/4}}\sqrt{\frac{N(t)}{\wdth_{x}(t)\wdth_{y}(t)\wdth_{z}(t)}}\,.
\end{equation*}%
Here we assumed that the center of mass of the solitary wave is at rest,
hence that the dynamics is restricted to the evolution of the longitudinal and
transverse widths. Besides the Gaussian widths, this ansatz contains several other real-valued variational parameters: $N(t)$, the instantaneous norm of the wave function; $\theta (t)$, the part of the complex phase that depends only on time but not on the spatial coordinates; and the so-called \dblqt{chirps}  $p_{x,y,z}$ \cite{Anderson1983_3135,Desaix1989_2441}. The necessity of these \dblqt{chirps} is well-known, but not really explained in the literature; we provide a brief discussion of them in the Appendix.

The GPE (\ref{3DGPE}) is the Euler-Lagrange equation that extremizes the action
 $S=\int_{V}dV\int dt\,\mathcal{L}$, with the Lagrangian density
\begin{gather*}
\mathcal{L}=i\frac{\hbar }{2}\left( \Psi \dot{\Psi}^{\ast }-\dot{\Psi}\Psi
^{\ast }\right) +\frac{\hbar ^{2}}{2m}\left\vert \nabla \Psi \right\vert ^{2}
\\
+\frac{1}{2}gN_{\text{a}}\left\vert \Psi \right\vert ^{4}+\frac{1}{2}m\left(
\omega _{r}^{2}\left( x^{2}+y^{2}\right) +\omega _{z}^{2}z^{2}\right)
\left\vert \Psi \right\vert ^{2}\,.
\end{gather*}%
To get a variational approximation, we substitute the Gaussian variational ansatz of Eq.~(\ref{ans}) into $\mathcal{L}$ and perform
the 3D spatial integration over all space. This results in an effective Lagrangian $L$ which is a function of the variational parameters. Now one demans that the usual Euler-Lagrange equations (ELEs) be satisfied for each variational parameter. The ELE for $\theta (t)$ gives that the norm $N(t)$ is constant in
time, and is therefore 1 at all times. The ELEs for the chirps enable us to express them in terms of the Gaussian widths: $p_{x}=m\dot{\wdth}_{x}/\wdth_{x}$, and similarly for $y$- and $z$-components [note that there is a typo in Ref.
\cite{Perez-Garcia1996_5320}, where $\hbar $ must not be squared in Eq.~(11a)]. Finally, the ELEs for the widths will depend also on the chirps, but not on $\theta (t)$. Since the chirps can be expressed in terms of the widths, we end up with a homogeneous system (\ref{VA}) involving the widths only and none of the other variational parameters. In principle, there is also an ELE for the norm, which gives the equation of motion for $\theta (t)$, but this is not of interest to us.

Clearly, there are essential features of the actual GPE dynamics that this
variational model cannot capture, such as the dwarf satellite peaks visible
at the half-period in Fig.~\ref{basic_breather} and at various points in
Fig.~\ref{complex_breather}, and the double-peak structure at $t=50$ in the
latter figure. Furthermore, the variational model does not allow for \emph{weak collapse} (in which only a vanishing fraction of the solitary wave ultimately reaches the sigularity), but rather only for \emph{strong} one (where that fraction is finite, in fact 100\% within the variatinal model). However, it is known that the 3D GPE only supports weak collapse \cite{Zakharov1975_465,Zakharov1985_154,Zakharov1986_773,Berge2002_136,Eigen2016_041058}. Nevertheless, the variational model helps to gain qualitative
understanding of the model's dynamics.

Equations (\ref{VA}) have the form of the Newton's equations of motion for a
particle in a 3D potential. To help with the analysis, we impose cylindrical
symmetry on the equations, setting $\wdth_{x}(t)=\wdth_{y}(t)$. Also, from now on we will work in the natural units (Appendix~\ref{app:natural}). To get the resulting
2D Newton's equations to also be derivable from a potential, we use a rescaled
radial coordinate, $\wdth_{x}(t)/\sqrt{2}=\wdth_{y}(t)/\sqrt{2}\equiv r(t)$, and
accordingly change the notation for the axial width, $\wdth_{z}(t)\equiv z(t)$. The appropriate 2D potential is then
\begin{gather}
V\left( r,\,z\,\left\vert \,\beta ,\,\gamma _{z}\right) \right. =  \notag \\
\frac{2}{r^{2}}+\frac{1}{2z^{2}}+\frac{\beta }{\sqrt{2}\pi ^{3/2}r^{2}z}+%
\frac{1}{2}\left( r^{2}+\gamma _{z}^{2}z^{2}\right) \,.  \label{var_pot_expr}
\end{gather}
Similar equations appear also in Ref.~\cite{Carr2002_063602}.

\begin{figure}[tbp]
\begin{center}
\includegraphics[width=0.5%
\textwidth,keepaspectratio=true,draft=false,clip=true]{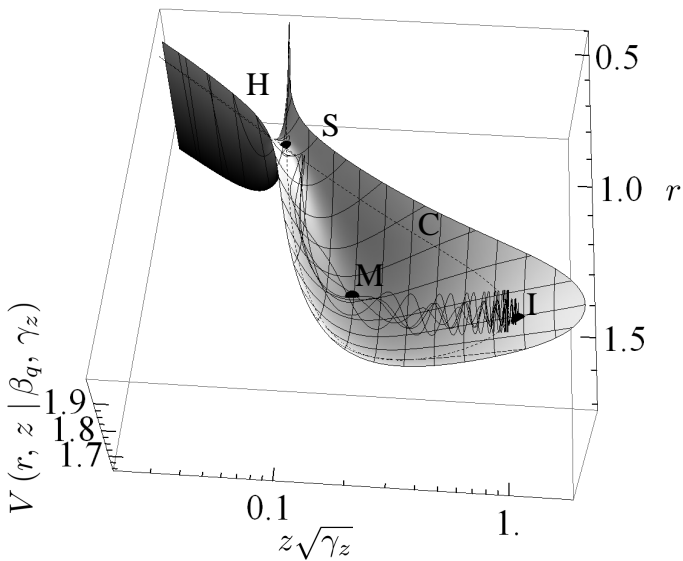}
\end{center}
\caption{ \textbf{The effective potential for the variational approximation.}
This is a plot of the potential $V\left( r,\,z\,\left\vert \,\protect%
\beta _{q},\,\protect\gamma _{z}\right) \right. $ in Eq.~(\protect\ref%
{var_pot_expr}) for $\protect\gamma _{z}=1/50$ and $\protect\beta _{q}=0.903%
\protect\beta _{\text{GS}}^{\text{V}}$. Here $\protect\beta _{\text{GS}}^{%
\text{V}}$ is the value of $\protect\beta $ at which potential $V$ ceases to
have a local minimum at this $\protect\gamma _{z}$ (i.e. $\protect\beta _{%
\text{GS}}^{\text{V}}$ is the variational prediction for the ground-state
critical value of the coupling, $\protect\beta _{\text{GS}}$). Point \textbf{%
M} is the local minimum of the potential, while \textbf{S} is the saddle
point. Region \textbf{H} marks the \dblqt{potential hole} which,
if the system ventures into it, leads to the collapse, driving both the
radial and axial widths, $r$ and $z$, to zero. Dashed curve \textbf{C} is
the equipotential level curve of the saddle point. The $(r,\,z)$ coordinates
of point \textbf{I} correspond to the minimum of the potential for $\protect%
\betai=0.0451\protect\beta _{\text{GS}}^{\text{V}}$. The system is first prepared in this minimum (i.e., in
its ground state for $\protect\beta =\protect\betai$), and then $%
\protect\beta $ is quenched by a factor of $q=20$ to $\protect\beta _{q}=0.903%
\protect\beta _{\text{GS}}^{\text{V}}$. Since the kinetic energy
is zero immediately after the quench, the post-quench potential energy at
\textbf{I} is the total energy of the effective particle, and, in the
particular case presented here, it will have enough energy to climb over the
saddle point, making the system unstable in the framework of the variational
model. For the system to be energetically protected against collapse (i.e.,
stable over infinite times) at this particular value of post-quench coupling
$\protect\beta _{q}$, the pre-quench coupling $\protect\betai$
should be at least $(1/8)\protect\beta _{\text{GS}}^{\text{V}}$, i.e., one
should limit the quench factor to $q<7.22$. Even so, as the plotted trajectory
shows, the system completes two full returns from the pericenter before
finally making it over the saddle point and collapsing. }
\label{var_pot}
\end{figure}

Thus, the \dblqt{breathing} of the matter wave is modeled by the
trajectory $(r(t),\,z(t))$ of a unit-mass classical particle in the potential $V$. A typical shape of this potential is shown in Fig.~\ref{var_pot}
(a similar surface was discussed in Ref.~\cite{Carr2002_063602}). For
given $\gamma _{z}>0$ and $\beta <0$, the potential has the following
features: if $|\beta |$ is not too large, there is a local minimum
(corresponding to the GS of the system, and labeled \textbf{M} in Fig.~\ref%
{var_pot}), a negative singularity at the origin (corresponding to the
\dblqt{black hole} \cite{Ueda1999_3317} that forms as a result of
the collapse, and labeled \textbf{H} in the figure), and a saddle point
between the two (labeled \textbf{S}). We always quench from the GS of some
potential; hence, the effective particle corresponding to the post-quench
state always starts from the rest position. If the initial position (labeled
\textbf{I}) belongs to the \dblqt{stability region}---defined as
the region that both (i) includes the local minimum and (ii) is bounded by
the saddle-point equipotential curve (labeled \textbf{C} in the
figure)---then the particle does not have enough energy to climb over the
saddle point and eventually fall into the singularity at the origin; hence,
the system is stable against collapse. On the other hand, if \textbf{I}
is outside the stability region, the collapse is energetically allowed.
However, the particle's trajectory may, initially at least, keep missing the
singularity. Intuitively, however, one may believe that this would not
continue forever: as the system is far from integrability, it should be
ergodic enough for almost every trajectory, whose energy is larger than the
value of potential $V$ at \textbf{S}, to pass, sooner or later, too close to
the origin and be sucked into the collapse singularity. Therefore, quenches
that place \textbf{I} outside the stability region always result in a system
that either collapses immediately, or is effectively metastable, collapsing
in a finite time. For example, in Fig.~\ref{var_pot}, we see a trajectory that misses the singularity twice before finally being sucked into it on the third approach; we found that small changes in the parameters can make a difference in which approach to the saddle point turns into a collapse.

\begin{figure}[tbp]
\begin{center}
\includegraphics[width=0.4%
\textwidth,keepaspectratio=true,draft=false,clip=true]{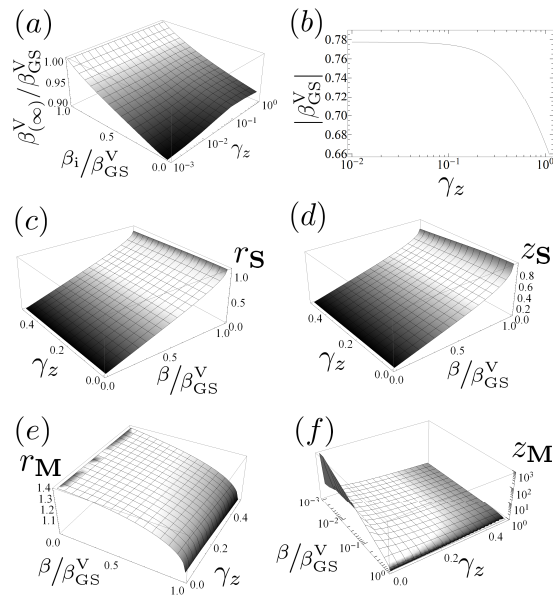}
\end{center}
\caption{ \textbf{Results from the variational model.} \textbf{(a)}
Within the variational theory, $\betacInfV$ is the strongest $\beta_{q}$ to which we can quench and still have the system be energetically protected from collapse. It is a variational estimate of $\betacInf$. Here we plot how $\betacInfV$ depends on the aspect ratio $\gamma_{z}$ and the pre-quench coupling $\betai$. \textbf{(b)}$\left|\betacGSV\right|$ is the least upper bound on $\left|\beta\right|$ such that potential $V$ in Eq.~(\ref{var_pot_expr}) still has a minimum. It is a function of $\gamma_{z}$, and is a variational estimate of $\left|\betacGS\right|$. Comparing these results to those in Table~\ref{GammalValues}, we see that the variational model overestimates $\left|\betacGS\right|$ by about 15\%.  In (c) we plot the $r
$-coordinate of the saddle point of the potential as a function of $\protect%
\beta $ and $\protect\gamma _{z}$. In (d)-(f), we do the same, respectively,
for the $z$-coordinate of the saddle point, and the $r$- and $z$-coordinates
of the potential minimum. }
\label{var_mod}
\end{figure}

As $|\beta |$ grows larger, the local minimum \textbf{M} and the saddle point
\textbf{S} become closer, until they coalesce at some critical value $\beta_{\text{GS}}^{\text{V}}$. For still larger values of $|\beta |$, the
potential contains no local minima and no saddle points, and the respective
particle is never energetically protected from falling into the singularity
(in particular, $\beta _{\text{GS}}^{\text{V}}$ is the variational estimate
of $\beta _{\text{GS}}$). Similarly to the reasoning above, the
non-integrability of the system then suggests that it will, with probability
$1$, collapse in a finite time, irrespective of the initial state. In this
case, the system may nevertheless be effectively metastable for some time, prior to the eventual onset of the collapse---at least if the
coupling strength is not much larger than its critical value. However, numerical integration of the variational equations of motion show that once the local minimum (and so the variational ground state) is gone, metastability is very weak. In contrast, it turns out that in the full GPE, metastability continues well beyond the point where the system has no stable ground state. This is why we will not quantitatively study metastability within the variational model. Instead, we will study two kinds of metastability below, using the full GPE. The informative aspects of the variational model, beyond the \dblqt{mental picture} of Fig.~\ref{var_pot}, will be presented in Fig.~\ref{var_mod}. First, however, we need to explain what functions will be plotted there.

We introduce some functions corresponding to the quantities discussed
above, namely $(r_{\text{\textbf{M}}}(\beta ,\,\gamma _{z}),\,z_{%
\text{\textbf{M}}}(\beta ,\,\gamma _{z}))$, $(r_{\text{\textbf{S}}}(\beta
,\,\gamma _{z}),\,z_{\text{\textbf{S}}}(\beta ,\,\gamma _{z}))$, and $\beta
_{\text{GS}}^{\text{V}}(\gamma _{z})$, the meaning of which is
self-explanatory. The superscript \dblqt{V} in the last item
implies that it is produced by the variational model, as opposed to $\beta _{%
\text{GS}}(\gamma _{z})$, whose values are generated by the full GPE.

Our procedure that models the quench dynamics is defined as follows. (1) Fix
a value of the aspect ratio, say $\gamma _{z,0}$, and of the coupling
constant, say $\betai$. The latter must be sub-critical, $%
\left\vert \betai\right\vert <\left\vert \beta _{\text{GS}}^{%
\text{V}}(\gamma _{z})\right\vert $, so that there is a stable GS for this
value of $\beta $. (2) Find the GS for these values of $\gamma _{z}$ and $%
\beta $, i.e., coordinates $(r_{\text{\textbf{M}}}(\beta _{\text{i}%
},\,\gamma _{z,0}),\,z_{\text{\textbf{M}}}(\betai,\,\gamma
_{z,0}))$ of the minimum of the potential $V\left( r,\,z\,\left\vert \,\betai,\,\gamma _{z,0}\right) \right. $. (3) Change the coupling
constant to $\beta _{q}=q\betai$, with quench factor $q>1$. In
order for the following steps to make sense, we must have $\left\vert \beta
_{q}\right\vert \leqslant \left\vert \beta _{\text{GS}}^{\text{V}}(\gamma
_{z})\right\vert $ (as mentioned above, in terms of the variational model,
the system can be, at best, metastable if $\left\vert \beta \right\vert
>\left\vert \beta _{\text{GS}}^{\text{V}}\right\vert $; below, we study
metastability using the full GPE). (4) Find coordinates $(r_{\text{\textbf{S}%
}}(\beta _{q},\,\gamma _{z,0}),\,z_{\text{\textbf{S}}}(\beta _{q},\,\gamma
_{z,0}))$ of the \emph{saddle point} of the post-quench potential, $V\left(
r,\,z\,\left\vert \,\beta _{q},\,\gamma _{z,0}\right) \right. $. Now we
define our long-time stability criterion in the framework of the variational
model: $V\left( r_{\text{\textbf{M}}}(\betai,\,\gamma
_{z,0}),\,z_{\text{\textbf{M}}}(\betai,\,\gamma
_{z,0})\,\left\vert \,\beta _{q},\,\gamma _{z,0}\right) \right. <V\left( r_{%
\text{\textbf{S}}}(\beta _{q},\,\gamma _{z,0}),\,z_{\text{\textbf{S}}}(\beta
_{q},\,\gamma _{z,0})\,\left\vert \,\beta _{q},\,\gamma _{z,0}\right)
\right. $ (note that the argument of $r_{\text{\textbf{M}}}$ and $z_{\text{%
\textbf{M}}}$ is $\betai$, while everywhere else it is $\beta
_{q} $). If this condition holds, then the effective particle (which,
post-quench, starts motion from the initial rest position) does not have
enough energy to climb over the saddle point and fall into the potential
hole. In this case, the variational model predicts that the system is stable
against collapse at all times. Accordingly, the \emph{critical} quenched
coupling strength $\beta _{(\infty )}^{\text{V}}(\beta ,\,\gamma _{z})$ is
defined as a solution of the equation $V\left( r_{\text{\textbf{M}}}(\beta
,\,\gamma _{z}),\,z_{\text{\textbf{M}}}(\beta ,\,\gamma _{z})\,\left\vert \,%
\smash\beta _{(\infty )}^{\text{V}},\,\gamma _{z}\right) \right. =V\left( r_{%
\text{\textbf{S}}}(\smash\beta _{(\infty )}^{\text{V}},\,\gamma _{z}),\,z_{%
\text{\textbf{S}}}(\smash\beta _{(\infty )}^{\text{V}},\,\gamma
_{z})\,\left\vert \,\smash\beta _{(\infty )}^{\text{V}},\,\gamma _{z}\right)
\right. $.

The results of the variational model are summarized in Fig.~\ref{var_mod}.
The conclusions are as follows. In Fig.~\ref{var_mod}(a), we see that the
maximal absolute value of the post-quench coupling strength (such that the
system is still stable against the collapse even for infinite times, in the
variational model) is about $90\%$ of $\beta _{\text{GS}}^{\text{V}}(\gamma
_{z})$ for small initial $\betai$, and it steadily increases up
to $100\%$ as $\betai$ approaches $\beta _{\text{GS}}^{\text{V}}$%
, while the dependence on $\gamma _{z}$ is weak. Figure~\ref{var_mod}(b)
implies that $\beta _{\text{GS}}^{\text{V}}$ overestimates the numerically
found value of $|\beta _{\text{GS}}|,$ which is given in Table~\ref%
{GammalValues}, by about $15\%$. Figures~\ref{var_mod}(c)-(f) show how $r_{%
\text{\textbf{S}}}$ and $z_{\text{\textbf{S}}}$, and $r_{\text{\textbf{M}}}$
and $z_{\text{\textbf{M}}}$, depend on $\beta $ and $\gamma _{z}$. Note
that, as $\beta $ approaches $\beta _{\text{GS}}^{\text{V}}(\gamma _{z})$, $%
(r_{\text{\textbf{S}}},\,z_{\text{\textbf{S}}})$ and $(r_{\text{\textbf{M}}%
},\,z_{\text{\textbf{M}}})$ approach a common value, which is close to $%
(1,\,1)$. Only $z_{\text{\textbf{M}}}$ has a strong dependence on $\gamma
_{z}$, and only for small values of $\gamma _{z}$ and $\beta $. In general, $%
r_{\text{\textbf{S}}}$ and $z_{\text{\textbf{S}}}$ are always roughly equal
to each other.  A simple explanation of this fact, which is arguably correct within the variational model, would go as follows: reaching the saddle point
heralds the beginning of the collapse, and, in the course of the collapse,
the attractive interactions are overwhelming everything else. Hence, the
shape of the collapsing condensate should be nearly spherical, at least within the variational model. Figures~\ref%
{var_mod}(c) and (d) suggest that $\beta /\beta _{\text{GS}}^{\text{V}}$ is
a good estimate of this approximately common value of $r_{\text{\textbf{S}}}$
and $z_{\text{\textbf{S}}}$.

We should note that the true asymptotic shape of a collapsing single-peak solution of the GPE has been extensively studied in the literature, and there is solid evidence that the collapse is indeed asymptotically isotropic in its final stages. However, there are also some results that contradict this claim, so that open questions still remain (Appendix~\ref{app:iso}).

\section{Numerical integration of the GPE \label{sec:numerics}}

The above variational model suggests that, even following a quench that
results in a supercritical coupling strength, the system may still avoid
collapse for some time, in a metastable state. By means of numerical
methods, we aim to address the following narrow but fundamental questions: (1)
What are the critical quenches that result in an \emph{immediate} collapse of
the system, i.e., without even a single return from the pericenter? (2) What
are the critical quenches that result in at least two returns from the
pericenter (i.e., one full cycle of the breathing)? (3) In the latter case,
as we approach the critical quench, what is the limiting value of the time
interval between the first two passages through the pericenter?

The simulations were run by means of the GPELab toolbox for MatLab \cite%
{Antoine2014_2969,Antoine2015_95}, which necessitates the use of a uniform
spatial grid. This makes it difficult to study large quenches, because the
system is initially broad (requiring a large computational domain), and then
drastically shrinks, requiring a closely spaced grid. For these reasons, we
here report the results for the pre-quench coupling strength $\geq
(1/8)\left|\beta _{\text{GS}}\right|$.

\subsection{Critical quench leading to the immediate collapse \label%
{subsec:immediate_collapse}}

For every $\gamma _{z}$ and every initial $\betai$, we steadily
increase the quench factor $q$ and keep track of $A_{\text{max,1}}$, the
peak height at the first post-quench pericenter. The collapse corresponds to
a divergence in $A_{\text{max,1}}$, and since the whole process looks like a
critical phenomenon, one may expect to see a power-law behavior. Indeed, we
always find that, as the quench factors approach the critical value, the
peak heights feature a power-law singularity:
\begin{equation}
A_{\text{max,1}}(\beta _{q})=\frac{a}{(b-\beta _{q})^{c}}\,,  \label{fit}
\end{equation}%
with $c>0$. We identify parameter $b$ with the $\beta $ corresponding to the
critical quench, $\beta _{(1)}$; see a typical example in Fig.~\ref%
{power-law_fig}.

\begin{figure}[tbp]
\begin{center}
\includegraphics[width=0.4%
\textwidth,keepaspectratio=true,draft=false,clip=true]{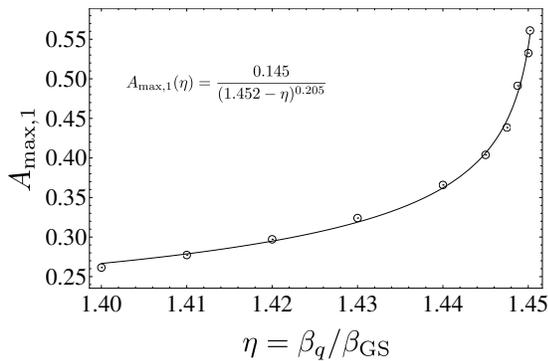}
\end{center}
\caption{\textbf{The power-law singularity for the first-pericenter peak
heights.} We start from the ground state for $\protect\gamma _{z}=1/1000$
and $\protect\betai=(1/8)\protect\beta _{\text{GS}}(\protect%
\gamma _{z})$, and then quench to $\protect\beta _{q}$. The open circles are the values of $A_{\text{max,1}}$, the peak height at the first post-quench
pericenter (for example, in Fig.~\protect\ref{complex_breather} this
corresponds to the peak height $\approx 0.16$, attained at $t=28$) The solid
line is a fit to the power law, whose analytical form is also displayed.
From the fit, we conclude that the critical value of $\protect\beta _{q}$ is
about $1.45\protect\beta _{\text{GS}}$. }
\label{power-law_fig}
\end{figure}

The results for $\beta _{(1)}$ obtained this way are summarized in Table~\ref{immediate_table}.
\begin{table}
\begin{center}
\textbf{
$\bm{\displaystyle \betacOne/\betacGS}$\\ as a function of $\bm{\displaystyle \betai/\betacGS}$ and $\bm{\displaystyle \gamma_{z}}$}\\
\mbox{}\\
\begin{tabular}{c |  d{1.3} d{1.3} c c c c}
\toprule
\begin{minipage}[c]{\widthof{\includegraphics[width=0.4in,clip=true,keepaspectratio=true,draft=false]{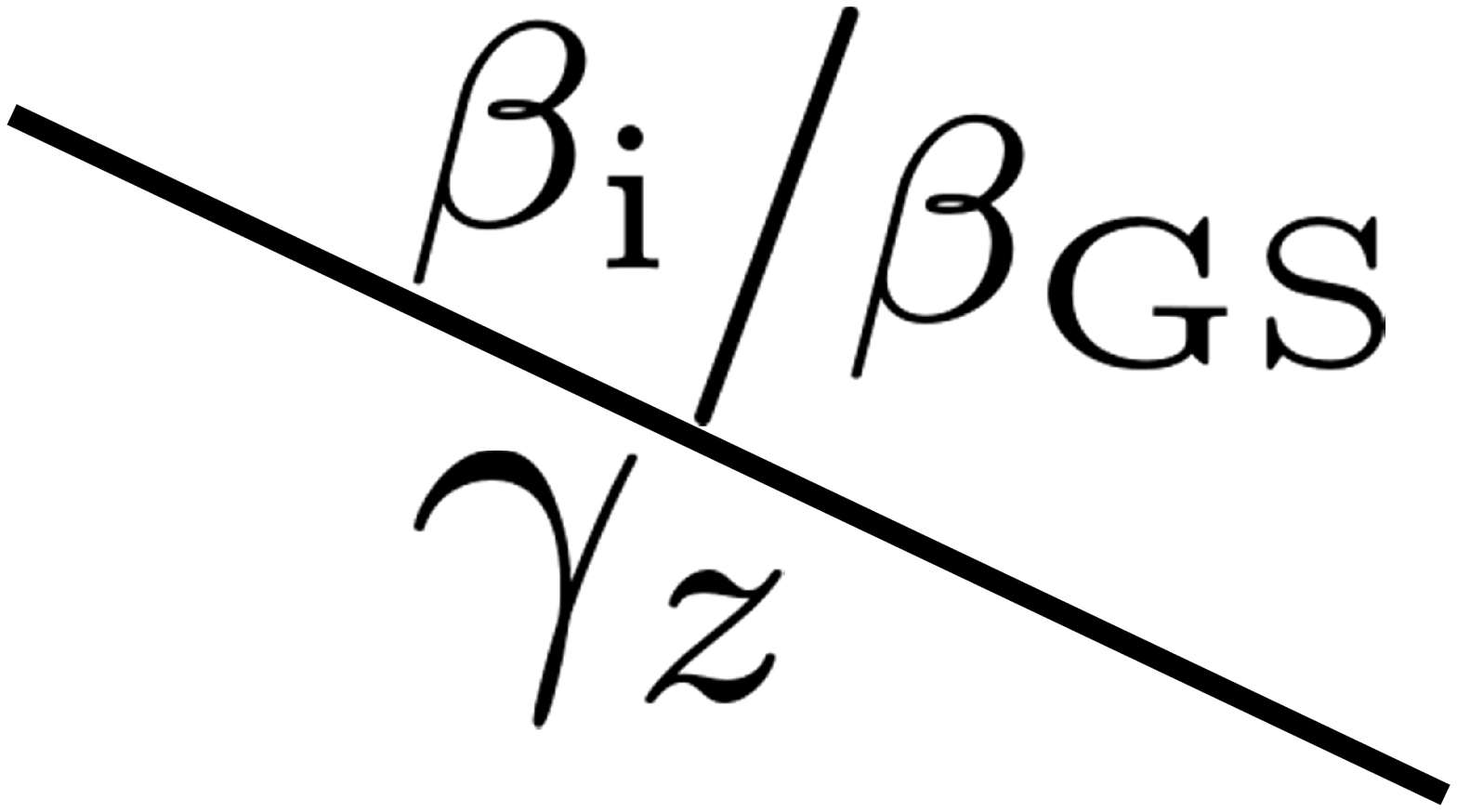}}}
\includegraphics[width=0.43in,clip=true,keepaspectratio=true,draft=false]{table_heading}
\end{minipage}
  & 0.125 & \mc{0.3} & 0.5 & 0.7 & 0.9  \\
\hline
\mbox{}\vspace{\blength}\mbox{} & & & & & \\
0 & 1.47  & 1.11 & 1.013 & 0.987 & 0.991  \\
1/1000 & 1.45 & 1.11 & 1.013 & 0.987 & 0.992  \\
1/300 & 1.34 & 1.11 & 1.013 & 0.987 & 0.992  \\
1/50 & 1.06 & 1.05 & 1.005 & 0.986 & 0.992
\\ \bottomrule
\end{tabular}
\end{center}
\caption{The values of $\betacOne\left(\betai,\,\gamma_{z}\right)/\betacGS\left(\gamma_{z}\right)$, where $\left|\betacOne\left(\betai,\,\gamma_{z}\right)\right|$ is the largest value of post-quench $\left|\beta\right|$ (i.e. of $\left|\beta_{q}\right|$) to which we can quench, starting from the ground state at $\betai$ and  $\gamma_{z}$, so that, prior to collapse, the breather performs at least one return from the pericenter. For example, starting from the ground state at $\gamma_{z}=1/300$ and $\betai=0.3\betacGS$ (where, according to Table~\ref{GammalValues}, $\betacGS=-0.676$), the system will immediately collapse if we quench to a $\left|\beta_{q}\right|>1.11 \left|\betacGS\right|$, whereas it will manage at least one return from the pericenter if $\left|\beta_{q}\right|<1.11 \left|\betacGS\right|$.}
\label{immediate_table}
\end{table}
\begin{table}
\begin{center}
\textbf{
$\bm{\displaystyle \betacTwo/\betacGS\left(\gamma_{z}\right)}$\\ as a function of $\bm{\displaystyle \betai/\betacGS}$ and $\bm{\displaystyle \gamma_{z}}$}\\
\mbox{}\\
\begin{tabular}{c |  d{1.3} d{1.3} c c c c}
\toprule
\begin{minipage}[c]{\widthof{\includegraphics[width=0.4in,clip=true,keepaspectratio=true,draft=false]{table_heading}}}
\includegraphics[width=0.43in,clip=true,keepaspectratio=true,draft=false]{table_heading}
\end{minipage}
 & 0.125 & \mc{0.3} & 0.5 & 0.7 & 0.9  \\
\hline
\mbox{}\vspace{\blength}\mbox{} & & & & & \\
0 & 1.233 & 1.076 & 1.013 & 0.987 &  0.991 \\
1/1000 & 1.217 & 1.075 & 1.016 & 0.987 &  0.992 \\
1/300 & 1.144  & 1.074 & 1.013 & 0.987  &  0.992 \\
1/50 & 1.031 & 1.032 & 1.005 & 0.986 & 0.992
\\ \bottomrule
\end{tabular}
\end{center}
\caption{The values of $\betacTwo\left(\betai,\,\gamma_{z}\right)/\betacGS\left(\gamma_{z}\right)$, where $\left|\betacTwo\left(\betai,\,\gamma_{z}\right)\right|$ is the largest value of the post-quench $\left|\beta\right|$ (i.e. of $\left|\beta_{q}\right|$) to which we can quench, starting from the ground state at $\betai$ and  $\gamma_{z}$, and still have the breather the breather completing at least two returns from the pericenter. For $\betai / \betacGS\geqslant 0.5$, the entries coincide with the corresponding ones in Table~\ref{immediate_table}, because it is the first approach to the pericenter that turns into a collapse (see text).}
\label{one_bounce_table}
\end{table}
\begin{table}
\begin{center}
\textbf{
Time interval between first two pericenters\\ as a function of $\bm{\displaystyle \betai/\betacGS}$ and $\bm{\displaystyle \gamma_{z}}$\\
in the limit as $\bm{\beta_{q}\to\betacTwo}$}\\
\mbox{}\\
\begin{tabular}{c |  c c c c c c}
\toprule
\begin{minipage}[c]{\widthof{\includegraphics[width=0.4in,clip=true,keepaspectratio=true,draft=false]{table_heading}}}
\includegraphics[width=0.43in,clip=true,keepaspectratio=true,draft=false]{table_heading}
\end{minipage}
& 0.125 & 0.3 & 0.5 & 0.7 & 0.9  \\
\hline
\mbox{}\vspace{\blength}\mbox{} & & & & & \\
0 & 72 & 29 & 41 & 26 & 20 \\
1/1000 & 71 & 30  & 29  & 38  & 27  \\
1/300 & 61 & 29 & 33 & 26 &  27 \\
1/50 & 31 & 28  & 27 & 26 & 23
\\ \bottomrule
\end{tabular}
\end{center}
\caption{The time intervals between the breather's first and second traversal of the breathing pericenter, in the limit as post-quench $\beta$ (i.e $\beta_{q}$) approaches $\betacTwo$. We quench from the ground state at $\betai$ and  $\gamma_{z}$.}
\label{one_bounce_times_table}
\end{table}
We see that the largest critical values of the coupling strength---almost $%
50\%$ larger in magnitude than the corresponding GS critical values $%
\left\vert \beta _{\text{GS}}(\gamma _{z})\right\vert $---occur for small
values of $\gamma _{z}$ and $\betai$. In other regimes, the
critical value of the post-quench $\beta $ is close to $\beta _{\text{GS}}$:
slightly larger than $\beta _{\text{GS}}$ if $\gamma _{z}$ is larger but $%
\betai$ is still small, and slightly smaller than $\beta _{\text{%
GS}}$ if $\betai$ is close to $\beta _{\text{GS}}$, regardless of
the value of $\gamma _{z}$. The lack of dependence on $\gamma _{z}$ when $%
|\beta |$ is large makes sense, because when $\betai$ is close to
$\beta _{\text{GS}}$, the GS has a very small size and a roughly spherical
shape, as in this case the interactions completely dominate over the
harmonic trapping.

\subsection{Critical quench leading to the delayed collapse \label%
{subsec:delayed_collapse}}

Here we seek another kind of critical quench, determined by the largest
value of the quench factor that still allows the system to return from the pericenter
at least twice. As the quench increases, the collapse that prevents the
system from accomplishing this may happen either on the second approach to
the pericenter, or already on the first. In the latter case, the critical
value of the post-quench coupling will be the same as in the previous
section. Numerically, we observe that the second approach leads to
collapse if the pre-quench coupling is weaker than about $\beta _{\text{i}%
}=(1/2)\beta _{\text{GS}}$, whereas if the pre-quench coupling is stronger
than that, then the system collapses on the first approach. Therefore, there
is a \emph{crossover} $\betai$ at which the occurrence of the
collapse switches from the second approach to the first. Thus we conclude
that, except for the crossover point $\betai$, for any given $%
\gamma _{z}$ and $\betai$, as the critical quench is approached,
the peak height increases, according to a singular power law, at exactly one
of the pericenters. The variational model of Sec.~\ref{sec:variational}
provides an intuitive reason for this: if a pericenter is a precursor of the
collapse (in the sense that it turns into a collapse if we slightly increase
the coupling strength), then the pericenter corresponds to the effective
particle approaching the origin in a somewhat fine-tuned, a
\dblqt{just-right} way. As the quench continuously increases, it
is not surprising that, eventually, \emph{one} of the first two visits by
the particle to the region around the origin starts to resemble the
\dblqt{just-right} way of the approach. But it would be
surprising if this started to happen for \emph{both} of the particle's first
two visits, the latter actually requiring another fine-tuning, this time in $%
\betai$, leading to the crossover $\betai$.

The critical value of the coupling can therefore be determined just as in
the previous section: by monitoring, as the quench increases, the peak
heights at the first two pericenters. Eventually, one of these heights will
start to increase (considered as a function of the quench) according to a
singular power law, and we can again use the curve-fitting method of Fig.~%
\ref{power-law_fig} to figure out the critical value of the coupling
strength. The corresponding results are summarized in Table~\ref%
{one_bounce_table}.

As the quench keeps increasing, one can also keep track of the time interval
between the first two pericenters. In the limit of the quench approaching the
critical point, these time intervals converge to finite values, which are
presented in Table~\ref{one_bounce_times_table}.

\section{Summary and outlook \label{sec:conclusion}}

We have numerically studied, within the mean-field (GPE-based)
approximation, the dynamical collapse of a class of excited states in a BEC
with attractive two-body interactions [characterized by the interaction
constant $\beta <0$ in Eq. (\ref{GPELabGPE})], in a cigar-shaped potential
trap of aspect ratio $\gamma _{z}$. The excited states in question are
produced by preparing the system in its GS (ground state) for some $(\gamma
_{z},\,\betai)$, and then quenching the coupling strength to some
$\beta _{q}$ with a larger absolute value. For relatively weak quenches, the
result is a stable excited state of the BEC that undergoes oscillations in
its width and amplitude, i.e., a breather. When the width of the system,
viewed as a function of time, attains a minimum or maximum, we say that the
system is at its \dblqt{pericenter} or \dblqt{apocenter,} respectively. On the other hand, for very strong quenches, the
system collapses immediately. For intermediate quenches, the system becomes
metastable, exhibiting several pericenter-apocenter breathing cycles before
collapsing. We have tabulated, for a range of values of $\gamma _{z}$ and $%
\betai$, the following quantities: (1) the smallest $\left\vert
\beta _{q}\right\vert $ starting from which the collapse occurs immediately
(Table~\ref{immediate_table}); (2) the smallest $\left\vert \beta
_{q}\right\vert $ at which the system collapses before completing two
returns from the pericenter, i.e., in the course of its first or second
shrinkage stage (Table~\ref{one_bounce_table}); (3) the limiting value, as $%
\left\vert \beta _{q}\right\vert $ keeps increasing towards the critical
value from Table~\ref{one_bounce_table}, of the time interval between the
first two pericenters (Table~\ref{one_bounce_times_table}). The collapse is
always identified by the following method: we monitor the peak
height of the density profile at the pericenters of interest, as $\left\vert
\beta _{q}\right\vert $ steadily increases. When $\beta _{q}$ gets close to
the critical value, the pericenter peak height features a singular
power-law dependence on $\beta _{q}$. By fitting the peak heights to the
power-law approximation given by Eq. (\ref{fit}), we have found the critical
value of $\beta _{q}$, see Fig.~\ref{power-law_fig}. In Sec.~\ref%
{sec:variational}, we have also studied the stability in the framework of
a variational model, which offers useful intuition for the understanding
of \mbox{(near-)collapse} dynamics. Moreover, this model provides the only
currently available estimate for the critical value of the quench below
which the system is stable for indefinitely long times [see Fig.~\ref{var_mod}(a)]:
ignoring the weak dependence on the aspect ratio $\gamma _{z}$, the value of
$\left|\beta\right|$ attained by the quench must fall below $\approx 90\%$ of $\left|\beta _{%
\text{GS}}(\gamma _{z})\right|$, for small initial $\betai$. This upper
limit increases approximately linearly towards $100\%$ of $\beta _{\text{GS}%
} $ as $\betai$ approaches $\beta _{\text{GS}}$.

As a possible extension of the present work, one can study what kinds of
parameter regimes allow for metastability on time scales at which the
upcoming experiments may run. If they may be much longer than a few
breathing cycles, or if the pre-quench interaction strength is weaker than $%
(1/8)\left\vert \beta _{\text{GS}}\right\vert $, the new theoretical studies
may require more powerful computational resources than what is used in the
present work.

\begin{figure}[tbp]
\mbox{}\newline
\par
\begin{center}
\includegraphics[width=0.4%
\textwidth,keepaspectratio=true,draft=false,clip=true]{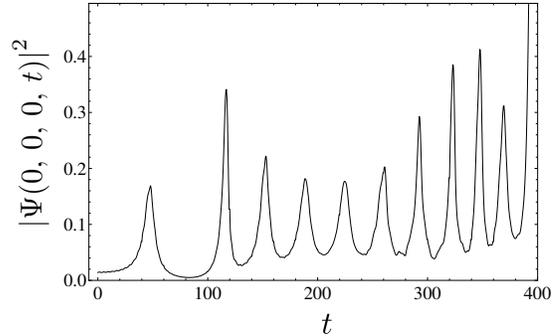}
\end{center}
\caption{ \textbf{Nonperiodic breathing.} The time evolution of
the density's peak height, following a quench from the ground state
corresponding to parameters $(\protect\gamma _{z}=0,\,\protect\beta _{\text{i%
}}=0.125\protect\beta _{\text{GS}})$ to $\protect\beta _{q}=1.23\protect%
\beta _{\text{GS}}$. The ten complete peaks correspond to pericenters; the 11th approach to the pericenter likely results in a collapse. The peak-to-peak time intervals go from $69$ for the first one, to $36$ for the second one, to $22$ for the ninth one (in the natural units).}
\label{nonperiodic_fig}
\end{figure}

On the other hand, if the time scales available to the simulations are
sufficiently long, we envision studying chaotic dynamics of
breathers. In particular, one can see in Fig.~\ref{nonperiodic_fig} that the
breathing need not be periodic.
The study of chaotic dynamics can be carried out at the level of
dynamics of classical particles \cite{Zaslavsky2005} (which actually
corresponds to the variational model), and at the level of the
classical-field dynamics \cite{Cvitanovic2000_61} (which corresponds to
simulations of the full GPE). Eventually, in this system one may also be
able to study quantum effects in chaos as well \cite{Wimberger2014}.

Next, an immediate objective may be to theoretically substantiate the
intuitively expected and numerically observed singular power-law dependence
of the pericenter peak heights on the coupling strength (Fig.~\ref%
{power-law_fig}). To the best of our knowledge, what has been extensively
studied so far are self-similar solutions that exhibit power-law divergences
as functions of \emph{time} for a fixed supercritical interaction strength
\cite{sulem1999}. In contrast, the power-law observed in Fig.~\ref%
{power-law_fig} is a function of the interaction strength, as it is
approaching the critical value from below.

Finally, for realistic description of experiments, one should include the loss and gain terms into the GPE. After all, when the BEC becomes unstable
and collapses, it loses particles through three-body
recombination. Thus closer the system comes to a collapse, the more relevent the loss terms are expected to be.

\begin{acknowledgments}
We thank R. G. Hulet and V. A. Yurovsky for valuable discussions, and the anonymous referee for alerting us to Refs.~\cite{Biasi2017_032216,Mardonov2015_043604} (as well as for very constructive comments). We
appreciate partial financial support from the National Science Foundation
grants PHY-1402249 and PHY-1607221, and from the Binational (US-Israel)
Science Foundation grant No. 2015616.
\end{acknowledgments}

\appendix

\section{The natural units \label{app:natural}}

The \dblqt{natural} units in this problem are those in which $\hbar=m=\omega_{r}=1$. It follows that the natural units of time, length, and mass, respectively, are $u_{T}=1/\omega _{r}$, $u_{L}=a_{\perp }=\sqrt{\hbar /(m\omega _{r})}$, and $u_{M}=m$.
In the natural units, the GPE simplifies to
\begin{gather}
i\partial_{t}\Psi=-\frac{1}{2}\nabla^2\Psi+4\pi\,\beta\left|\Psi\right|^{2}\Psi \notag \\
+\frac{1}{2}\left(x^{2}+ y^{2}+\gamma_{z}^{2} z^{2}\right)\Psi\,,
\label{GPELabGPE}
\end{gather}
where we used the fact that, in the natural units, $\omega_{z}$ has the numerical value $\gamma_{z}$, and $g N_{\text{a}}$ has the numerical value $4\pi\,\beta$. Of course, the above equation can also be interpreted as a change of variables: we introduce $\bar{x}=x/u_{L}$ (and similarly for the $y$- and $z$-components) and  $\bar{t}=t/u_{T}$, and define the function $\bar{\Psi}(\bar{x},\,\bar{y},\,\bar{z},\,\bar{t})=u_{L}^{3/2}\,\Psi(u_{L}\bar{x},\,u_{L}\bar{y},\,u_{L}\bar{z},\,u_{T}\bar{t})$. Note that $\bar{\Psi}$ is normalized to 1 when integrated over $\bar{x}$, $\bar{y}$, and $\bar{z}$. We invert this, obtaining $\Psi(x,\,y,\,z,\,t)=u_{L}^{-3/2}\,\bar{\Psi}(x/u_{L},\,y/u_{L},\,z/u_{L},\,t/u_{T})$, and insert it into Eq.~(\ref{3DGPE}). We then express $x$, $y$, $z$, and $t$  in terms of $\bar{x}$, $\bar{y}$, $\bar{z}$, and $\bar{t}$, and simplify. We get a \dblqt{barred} version of Eq.~(\ref{GPELabGPE}): that very equation  except that all of $x$, $y$, $z$, $t$, and $\Psi$ have bars on them. But since $\bar{x}$ is the numerical value of $x$ when using natural units (and similarly for $y$, $z$, $t$ and $\Psi$), we see that the \dblqt{barred} version of Eq.~(\ref{GPELabGPE}) is numerically identical to the actual Eq.~(\ref{GPELabGPE}) \emph{provided} that in the latter we use the natural units. And we have already seen that in those units, Eq.~(\ref{GPELabGPE}) is numerically identical to Eq.~(\ref{3DGPE}).

\section{The \dblqt{chirps} in the variational ansatz \label{app:chirps}}

It is well known that any variational ansatz for the wave function $\Psi$, like that in Eq.~(\ref{ans}), must have a spatially-dependent phase whose functional form is $x^2$ times a function of time, plus the corresponding terms for the $y$- and $z$-components. (For example, in 1D, the functional form of the phase would be the same if instead of the Gaussian density profile one assumed a $\sech$-squared \cite{Desaix1991_2082}, which is the exact profile of a single GPE soliton.) Despite it being well-known, it has proved difficult to find a published reference that explains these facts. Reference~\cite{Perez-Garcia1996_5320} says that Ref.~\cite{Desaix1991_2082} has show that these terms `are essential if one wants to obtain reliable results'. From this one might expect that the latter reference compares what happens when these terms are excluded as compared to when they are present. But this is not what that reference does: instead, it compares the variational method (where such a term is simply included without comment) to a different, non-variational method. Indeed, we can see that completely excluding these terms leads to a stationary solution. Namely, excluding these terms is equivalent to setting $p_{x}=0$ (and similarly for the other two components); but since $p_{x}=m\dot{\wdth}_{x}/\wdth_{x}$, we get $\dot{\wdth}_{x}=0$ (and similarly for the $x$- and $y$-components).

The closest to an explanation that we've been able to find in primary literature appears in Ref.~\cite{Desaix1989_2441}, which says only that these chirps should expected from `[p]hysical intuition based on well-known results from linear and nonlinear pulse propagation theory'. On the other hand, Shaw \cite{Shaw2004} suggests that these terms were motivated by the exact solution for a freely expanding Gaussan, a point to which we will return shortly.

Moreover, the literature seems to be in agreement \cite{Shaw2004,Mikhailov1999} that Ref.~\cite{Anderson1983_3135} is the original reference for this functional form; however, in that paper it just appears without comment.

Here we will try to justify the presence and the functional form of these chirps in the Gaussan ansatz a bit more explicitly. First of all, the exact solution for a freely propagating Gaussian wavepacket with a stationary center of mass has this form \cite{Shaw2004}. Second, one can always write $\Psi=\sqrt{\rho} e^{i R/\hbar}$, where $\rho$ and $R$ are real-valued functions of $x$, $y$, $z$, and $t$. Inserting this into the GPE, the imaginary part of the resulting equation gives the equation of continuity, $\dot{\rho}+\vec{\nabla}\cdot\left(\rho \vec{\nabla}R/m\right)=0$. And from the latter it immediately follows that as soon as the density $\rho$ has any time dependence at all, the phase $R$ must have a nonzero spatial gradient and so must depend on the spatial coordinates. Moreover, in 1D it is easy to show that if $\rho$ is a Gaussian with width $\wdth_{x}(t)$, then in fact the only way to satisfy the 1D continuity equation is to have $R=\frac{1}{2}x^{2}m \dot{\wdth}_{x}/\wdth_{x}$. The ansatz in Eq.~(\ref{ans}) is a natural generalization of this, and it is easy to show that it does satisfy the 3D equation of continuity provided that $p_{y}=m \dot{\wdth}_{y}/\wdth_{y}$, and similarly for the $z$-components. Admittedly, we don't have a proof that this functional form of the phase is the only possible one that will satisfy the 3D continuity equation when $\rho$ is a 3D gaussian, but it is a reasonable conjecture. Finally, this ansatz results in Eqs.~(\ref{VA}), and they are known to give very good results when compared to numerics---for example for the frequencies of small oscillations around the equilibrium point \cite{Perez-Garcia1996_5320}, and for predictions for the onset of collapse, equilibrium widths, and dynamical evolution laws of the condensate parameters \cite{Perez-Garcia1997_1424}.

We hope that all these considerations provide a sufficient justification for the presence of the \dblqt{chirps} in the variational ansatz.

\section{Asymptotic isotropy of a collapsing single-peak solution of the GPE \label{app:iso}}

The true asymptotic shape of a collapsing single-peak solution of the GPE has been extensively studied in the literature, but open questions and even a degree of controversy still remain. On the one hand, it is known that asymptotically anisotropic collapsing solutions do formally exist  \cite{Pelletier1987_187} and have been reported in earlier numerically studies \cite{Degtyarev1974_365}. Moreover, these works gave physical arguments why it is that the anisotropic rather than isotropic collapsing solutions should be stable. On the other hand, not only has a numerical linear stability analysis showed that isotropic collapse is stable with respect to anisotropic disturbances \cite{Vlasov1989_1945}, but also the most recent high-quality numerical studies have never seen any asymptotically anisotropic collapsing solutions. Instead they have found that initially anisotropic one-peak solutions become isotropic near the collapse point, see Refs.~\cite{Landman1991_393,Akrivis2003_186} and p.~130 of Ref.~\cite{sulem1999}. Unfortunately, analytic results are still lacking.

\section{Discussion of results of Biasi et al. \cite{Biasi2017_032216} in light of our own results \label{sec:Biasi}}

Biasi et al. \cite{Biasi2017_032216} carried out a study very much related to ours, but with the following differences: 1. their system was radially symmetric, while ours is not; 2. their initial states were Gaussians, whereas for us they were the ground states of our system; 3. the number of spatial dimensions they considered went from two all the way to seven, whereas we only studied the 3D case; 4. their system was propagated until either it collapsed or some a priori specified maximal time was reached, no matter how many times it \dblqt{bounced} from the pericenter; in our case, we concentrated on thresholds for transition between immediate collapse and collapse after a single bounce, and for transition between collapse after a single bounce and collapse after two. For some parameter choices their system was at these same transition points (in terms of how many bounces the system was able to complete before collapsing), and so in those particular cases (e.g. those labeled $\epsilon_{1}$ and $\epsilon_{2}$ in their Fig.~1), their system and ours do not differ with respect to the number of completed bounces; 5. they studied temporal evolution of the energy spectrum of low-lying modes as well as the mode-mode coupling coefficients, whereas we have not; 6. their criterion for collapse was apparent numerical divergence of the height of the central peak of the density distribution, whereas we used the procedure outlined in Sec.~\ref{subsec:immediate_collapse}.

The plots for $\epsilon_{1}$ and $\epsilon_{2}$ in their Fig.~1 are consistent with the general pattern we observed in our data, and they would correspond to $\betacOne/\betacGS=0.67514$ (see our Table~\ref{immediate_table}) and $\betacTwo/\betacGS=0.67510$ (see our Table~\ref{one_bounce_table}), respectively. In our terms, we have that post-quench $\beta=\frac{1}{8}\sqrt{\frac{\pi}{2}}\epsilon^{2}\sigma^{3}$, where $\sigma$ is $\sqrt{2/5}$ in their Fig~1 and $\sqrt{1/2}$ in their Fig~2. Of course, since their initial states are not the same as ours, we wouldn't be able to directly compare the results quantitatively even if we had done simulations for $\gamma_{z}=1$.

Their Fig.~2 shows that the total time to collapse is a very complicated non-monotonic function of the coupling strength $\epsilon$. We don't have a plot corresponding to it, but we can make an educated guess as to what is happening. First of all, note that both the number of bounces and the time intervals between them generally change with changing $\epsilon$, which could easily result in a complicated plot. The most striking feature of their Fig.~2, however, are the sharp peaks in the time-to-collapse. These are likely the result of the system \dblqt{lingering} at the threshold of collapse for that particular bounce, a bit like when an object climbs a hill with just barely enough energy to go over it. (As we have seen above, variationally speaking, actually it is not merely the question of total available energy, but also of the correct \dblqt{aim} for the saddle point; see Sec.~\ref{sec:variational}.) And once $\epsilon$  passes (as it is lowered) this tricky threshold value, the collapses may start happening more quickly---until the next time $\epsilon$ is fine-tuned at the threshold of collapse during a particular bounce, which would produce another peak in their Fig.~2.  The scenario we just outlined is consistent with their statement, `Roughly speaking, each step corresponds to a number of bounces in the harmonic potential. At the boundary between steps, $t_{c}$ presents a bump'. Unfortunately, we cannot confirm this scenario from our own data, because in effect we have the equivalent of just two points from their Fig.~2. Note that what is being varied in our Tables~\ref{immediate_table}) and \ref{one_bounce_table} are the aspect ratio and the initial state, both of which are completely fixed in their Fig.~2; if we fix them, then each table gives just one threshold value of post-quench $\beta$.

%

\end{document}